\documentstyle[prl,aps]{revtex}

\begin{document}
\date{\today}
\draft
\twocolumn
\input epsf

\title{Ballistic Electron Emission Microscopy 
on CoSi${}_2$/Si(111) interfaces:\\
band structure induced atomic-scale resolution 
and role of localized surface states}

\author{K. Reuter$^{1,3}$, F.J. Garcia-Vidal$^2$, P.L. de Andres$^1$, 
F. Flores$^2$, and K. Heinz$^3$}

\address{$^1$Instituto de Ciencia de Materiales (CSIC), 
Cantoblanco, E-28049 Madrid (Spain)}

\address{$^2$Departamento de F\'{\i}sica Te\'orica de la Materia 
Condensada (UAM) 
E-28049 Madrid (Spain)} 

\address{
$^3$Lehrstuhl f\"ur Festk\"orperphysik,
Universit\"at Erlangen-N\"urnberg, Staudtstr. 7, 
91058 Erlangen (Germany)}

\maketitle

\begin{abstract}
Applying a Keldysh Green`s function method it is shown that hot electrons
injected from a STM-tip into a CoSi${}_2$/Si(111) system form a highly
focused beam due to the silicide band structure. This explains the
atomic resolution obtained in recent Ballistic Electron Emission Microscopy
(BEEM) experiments. 
Localized surface states in the $(2 \times 1)$-reconstruction
are found to be responsible for the also reported anticorrugation
of the BEEM current. These results clearly demonstrate the importance
of bulk and surface band structure effects for a detailed understanding of BEEM data.

\end{abstract} 


Ballistic Electron Emission Microscopy (BEEM),
and its spectroscopic counterpart (BEES), are powerful
techniques invented for exploring the electronic properties 
of metal-semiconductor (M-S) interfaces\cite{kaiser}.
Thin metallic films are deposited on different semiconductor materials
and the BEEM current, i.e. the current arriving at the
semiconductor after injection into the metal surface
from a Scanning Tunneling Microscope (STM) tip,
is measured as a function of the tip-metal voltage\cite{mario}.
The interpretation of these experiments is based on a three-step model: 
(i) first, electrons are injected from the tip into the metal (tunneling); 
(ii) then, electrons propagate through the film suffering collisions with 
different quasiparticles (transport), and 
(iii) finally, electrons overcome the Schottky barrier and enter into 
the semiconductor (matching of metal and semiconductor wavefunctions across the interface). 
The difficulty in analyzing experimental BEEM data stems 
from the strong influence
of all three steps requiring a careful theoretical modeling to
avoid spurious correlations between the parameters involved. Recently, 
it has been shown that the electronic band structure of the metal, which had
been completely neglected in earlier free electron models, plays a crucial
role in this regards \cite{prl}.

Recent experimental BEEM investigations on metallic silicide films deposited
on Si show (i) an atomic scale resolution of the M-S interface\cite{kanel1},
and (ii) a striking dependence of the interface BEEM current on the silicide
surface topography\cite {kanel2}. 
Dislocations and point defects at the interface
were well visible giving direct access to its quality. This is rather important
in view of the interface`s role in building the Schottky barrier or with
respect to the growth mode of silicides which are promising materials for
microelectronic applications\cite{scirep}. Equally important is the
quantitative understanding of how the obtained atomic scale resolution is
produced and why the BEEM current is related to the surface topography.
In this paper, we show that the high lateral resolution is caused by the
silicide`s band structure, which in the case of CoSi${}_2$/Si(111), on
which we concentrate, makes the electrons focus in the $<\!\!\!111\!\!\!>$
direction. This tells that the experimentally observed focusing is an
intrinsic feature of such films that might be exploited in future
applications. Additionally, the introduction of the appropriate surface
electronic structure explains the BEEM current dependence on the tip
position mainly as a result of the weight of localized surface states
on the reconstructed surface.

We use a full quantum-mechanical description of the BEEM problem based
on a Keldysh Green's function method \cite{prl}. This formalism presents
the important advantage over standard E-space Monte-Carlo approaches of 
yielding an appropriate description of the electronic band structure.
Moreover,
inelastic effects associated
with electron-electron interactions are also included in our method by adding
a positive imaginary part to the energy of the electron.
In order to analyze the
first two steps of the BEEM process, we choose a local orbital basis
for the description of the electronic structure of the tip and sample and the 
coupling between them. 
In particular, for CoSi$_2$ we use a slight modification 
of the tight-binding parameters
given in \cite{sanguinetti}, that accurately reproduce the band structure
of this silicide around the Fermi level \cite{mattheiss}. 
For the analysis of the interaction between tip
and sample we assume that only the last atom in the tip (0) is connected to the sample.
Hence, we express the coupling tip-sample in terms of a set of hopping matrices
$\hat{T}_{0 m}$, that link the tip atom (0) with the atom ($m$) in the sample surface. 
For each $\hat{T}_{0 m}$, a WKB derived exponential damping is applied, 
valid because the tip-sample distance in BEEM is rather large.

Being interested in understanding the observed nanometric spatial \
resolution of this
technique, we first analyze currents in real space. Within our formalism, 
the current between two sites 
$i$ and $j$ in the metal can be obtained from the
following formula \cite{prl,prb}:

\begin{eqnarray}
J_{ij}(V) = \frac{4e}{\hbar} \Im \int_{eV_o}^{eV} \!\! Tr \! \sum_{m n}
\left[ \hat T_{ij} \hat g_{jm}^{R} \hat T_{m 0} \hat \rho_{00}
\hat T_{0 n} \hat g_{ni}^{A} \right] dE,
\label{jreal}
\end{eqnarray}

\noindent
where $\hat T_{ij}$ 
is the hopping matrix linking local orbitals of both sites
($i$ and $j$), and the trace denotes summation over these orbitals.
$\hat g_{jm}^R(E)$ is the retarded Green's function for the surface
decoupled from the tip. This function
describes the propagation of an electron between atoms 
$j$ and $m$ inside the metal,
including the effect of the surface.
Atom $m$ in Eq. (1) is coupled to the tip atom $0$ by a hopping
matrix $\hat T_{m0}$, 
and $\hat \rho_{00}(E)$ is the density of states matrix at the
tip atom. The advanced Green's function, $\hat g_{ni}^A(E)$,
describes the electron propagation from an
atom $n$ at the surface down to the atom $i$, closing the loop
to give the current 
between atoms $j$ and $i$. The summation runs over all
tunneling active atoms 
in the sample surface, $m$ and $n$. The energy integration
is performed between the Schottky barrier ($eV_o$, assumed to be 0.66 eV
\cite{kanel1}) and the applied voltage ($eV$). 
However, due to the exponential
energy dependence of the coupling matrices, 
$T_{m0},T_{0n}$, the integrand is
a strongly increasing function with energy, 
so that already the contribution
at the highest energy ($eV$) 
provides the dominant fraction of the elastic BEEM
current in the near threshold region. 
To elucidate the physics behind the
observed effects, the presented results will therefore be restricted to
this highest energy.

\begin{figure}
\epsfxsize=0.5\textwidth \epsfbox{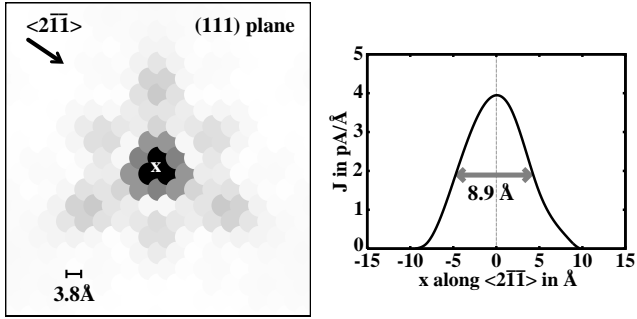}
\caption{Current distribution in a Si2 layer parallel to the
surface after propagation through 30{\AA} CoSi${}_2$(111) film. 
Injection from
the tip at $1.5eV$ occured in the center of the shown plane (white X),
where the maximum current propagating in a focused beam along the
$<\!\!111\!\!\!>$ direction can still be found. 
The linear gray scale indicates
current intensity at each atomic site:
black maximum to white zero current. The right hand
panel displays a cut through the focused beam 
in $<\!\!2\bar{1}\bar{1}\!\!>$
direction from which a FWHM of 8.9{\AA} can be derived.}
\label{1x1real}
\end{figure}

With Eq. (\ref{jreal}), 
the elastic propagation of electrons in real space
from the tip down 
to the M-S interface can be followed. In order to obtain
the final BEEM current, 
we further need to calculate the momentum distribution
of the electrons 
that reach the M-S interface, $J_{I}(E,{\bf k_{\parallel}})$.
This momentum distribution can be expressed as \cite{prb}:

\begin{equation}
J_{I} (E,{\bf k_{\parallel}}) = \frac{4e}{\hbar} \Im \; Tr  \sum_{b}
\left[ \hat T_{bc} \hat g_{c1}^{R} 
\hat T_{10} \hat \rho_{00} \hat T_{01} \hat g_{1b}^{A} \right],
\label{jk}
\end{equation}

\noindent
where in this case 
$\hat g_{c1}^{R}(E,{\bf k_{\parallel}})$ is the retarded
Green's function for the unperturbed metal, 
linking layer $c$ (the metal layer
at the M-S interface) and the surface 
layer $1$ which is connected to the tip
by a hopping matrix 
$\hat T_{10}({\bf k_{\parallel}})$. $\hat T_{bc}({\bf k_{\parallel}})$
is the hopping matrix connecting all 
upper layers $b$ with the interface layer $c$ and
finally $\hat g_{1b}^{A}(E,{\bf k_{\parallel}})$ 
is the advanced Green's function
linking the surface layer with layer $b$. 
These advanced and retarded Green's
functions and the ones appearing in 
Eq. (\ref{jreal}) can be readily computed using
renormalization group techniques \cite{guinea}.

\begin{figure}
\epsfxsize=0.5\textwidth \epsfbox{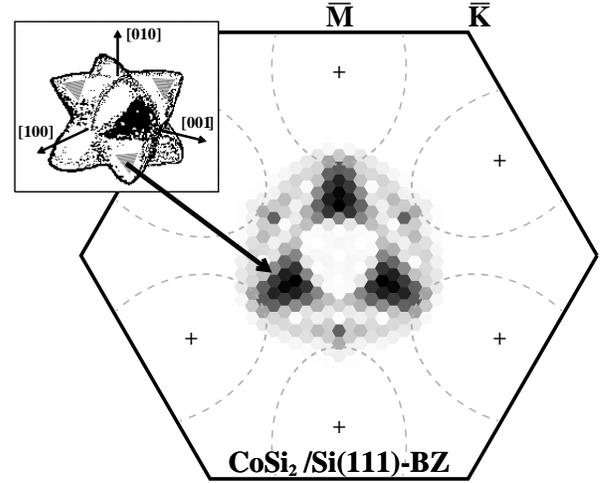}
\caption{Electronic current distribution 
in the 2D interface Brillouin zone, 
$J_{I} (E,{\bf k_{\parallel}})$, evaluated at $1.5$eV after 30{\AA} film
propagation. The current intensity is 
drawn with a linear gray scale, black representing
maximum current. Also shown are the ellipsoids defining available 
states in the semiconductor below $1.5$ eV.
The inset contains the constant energy surface
sheet mainly responsible 
for the current propagation: the shaded flat terraces
point in $<\!\!111\!\!\!>$ 
direction and correspond to the dark areas of the 2D current
distribution.}
\label{1x1k}
\end{figure}

First of all, 
we apply our formalism to the case of a CoSi$_2$(111)-(1x1)Si-rich
surface terminated 
film \cite{starke} and analyze the propagation of electrons 
from the tip to the CoSi${}_2$/Si(111) interface in real space. 
In the $<\!\!\!111\!\!\!>$
direction, this metal may be characterized by a 
stacking sequence of Si1-Co-Si2
trilayers (cf. Fig. 3a) 
with the interface to Si mainly formed below a Si2-type layer
\cite{starke}. Fig. \ref{1x1real} shows the current distribution
on every atom in such a Si2 
layer 30{\AA} below the surface as to compare
with experiments performed on films of equal width\cite{kanel1}.
The prominent effect we deduce from this figure is that the electrons
injected into the silicide are focused inside a 
very narrow beam propagating
perpendicular to the film. 
The right hand panel in Fig. 1 shows the intensity
in real space along a line 
in $<\!\!2\bar{1}\bar{1}\!\!>$ direction through the
center of the beam: 
the obtained FWHM of 8.9{\AA} compares very well with the
resolution of $\approx10.0${\AA} with which interface point defects
could be resolved experimentally 
in such films \cite{kanel1}. This, up to now,
highest achieved spatial 
resolution with BEEM had been impossible to explain 
assuming free electron propagation inside the metal,
predicting beam widths of 25 {\AA} for the same distance.

The electron focalization 
is due to the particular shape of the constant energy
surface sheet responsible 
for the major current propagation (see inset of Fig. \ref{1x1k});
it can be shown that between $E_F$ and $E_F + 2.5$ eV 
these sheets are practically the same except by 
a uniform shrinkage that increases linearly with energy. 
The shaded regions are nearly flat terraces perpendicular to the
$<\!\!\!111\!\!\!>$ direction,  
and act as a kind of
``condenser lens'' on the electron beam, 
keeping the electrons with corresponding
${\bf k}$-vector propagating 
along the $<\!\!\!111\!\!\!>$ direction \cite{prl}. This reasoning is
complementary to the current distribution we have calculated in
${\bf k}_{\parallel}$-space 
using Eq. 2 and shown in Fig. \ref{1x1k}. The three dark regions of
the 2D Brillouin zone where 
the ${\bf k}_{\parallel}$-current is mainly concentrated
correspond to the flat areas of the constant energy surface. 

The onset of BEES I(V) characteristics
is linked to the Schottky barrier height between
the metal and the semiconductor.
In our calculations 
that onset appears at $0.9$eV, $0.24$eV larger than the
Schottky barrier height commonly accepted for the 
CoSi$_{2}$-Si interface\cite{fotoemision,sirringhaus96}.
This is related to the assumed ${\bf k}_{\parallel}$-conservation
and to the absence of states in the metal matching the
conduction band minima in the semiconductor. 
The same delayed onset has been 
obtained by Stiles and Hamann\cite{stiles},
and it has been argued\cite{sirringhaus96,davis}
that a smaller onset can appear if a non
${\bf k}_{\parallel}$-conserving scattering process is operative for the
injected electrons at the silicide-silicon interface.
Indeed, the results reported by these authors for CoSi${}_2$/Si(111)
seem to point out that the effect of such processes
is to modify only slightly the BEEM-current beyond $0.9$ eV, 
but is enough to yield the appropriate M-S barrier height
at $\approx 0.66$ eV.
Therefore, for energies larger than $0.9$ eV,
current injection conserving ${\bf k}_{\parallel}$ dominates
the spectra, as expected intuitively from the good matching
between the Si and CoSi$_{2}$ lattices, and our theory applies.
We should also mention, 
however, that non ${\bf k}_{\parallel}$-conserving 
processes must play an important role for the 
BEEM contrast of defect images at the M-S interface due to the 
nanometric size of the electron beam.

In our next step we consider the case of the $(2 \times 1)$
surface structure of CoSi$_{2}$/Si(111).
Stalder et al.\cite{stalder}
reported this Si-rich reconstruction for
CoSi$_{2}$(111) with a geometry very similar to
Pandey's $\pi$-bonded chain model\cite{pandey}.
Fig. 3a shows a sideview of this surface
geometry with its topmost Si-bilayer reconstructed in
alternating high and low chains.
We have analyzed how the
geometry of the reconstructed $(2 \times 1)$
surface modifies the electron focalization
discussed above for the $(1 \times 1)$ 
surface. 
The main effect of the reconstruction is to
broaden the FWHM of the focused beam to
$13.6${\AA}.
This effect, that we associate to a larger
area of the surface unit cell where the
tunneling electrons are injected, has also
been observed experimentally by Sirringhaus
et al\cite{kanel1}.

A very interesting result observed for this reconstruction is 
that in the constant-current STM mode, the BEEM image of the
interface reflects the atomic surface periodicity, but out 
of phase with the topographic corrugation \cite{kanel2}.
This BEEM anticorrugation has previously been attributed to
atomic-scale variations 
of the energy tunneling distribution of the injected 
electrons \cite{kanel2}.
In order to analyze these results, 
we calculate
the current that reaches the M-S
interface as a function of the tip position
for a constant tunneling current ($1$ nA).
In general, to compute the BEEM current
injected into the semiconductor we would
need to use a transmission coefficient,
$T(E,{\bf k}_{\parallel})$, determined
from the matching of states at the interface.
However, this is not necessary to study the
particular dependence of the BEEM current
on the tip position, as $T$ is independent
of the tunneling injection. Moreover, we have
found that the ${\bf k}_{\parallel}$ distribution of the
current is nearly the same for
all the different positions of the tip, 
in accordance with the conclusions raised in \cite{kanel2}.  
Therefore, to study the effect that the
surface reconstruction introduces in the BEEM
current we can simply analyze the total 
current reaching the interface.
Fig. 3b shows that this quantity 
presents anticorrugation with respect to
the one found in the surface reconstruction.

\begin{figure}
\epsfxsize=0.5\textwidth \epsfbox{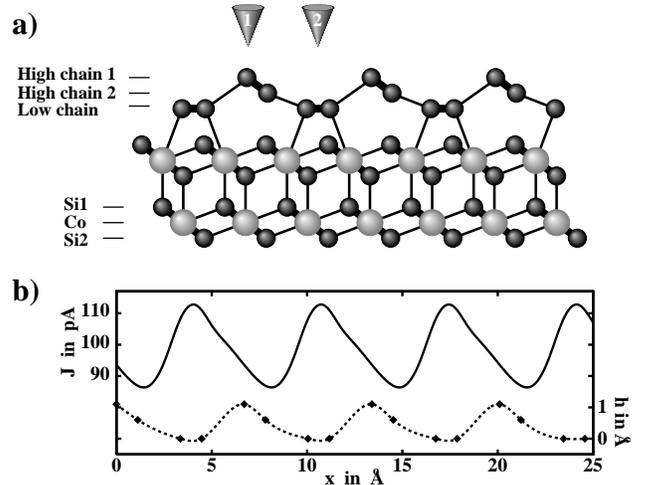}
\caption{
a) Sideview of the ($2 \times 1$)
surface reconstruction with tip positions 1
(high chain injection) and 2 (low chain injection).
b) M-S interface current at $1.5$eV after 30{\AA} film
propagation as a function of the tip position (higher curve,
left scale in pA). The lower curve gives a schematic surface
topography along the scan line (right scale in {\AA}).}
\label{anticorrugation}
\end{figure}

To understand the physics behind
this effect, we have studied the injected
current along the different metal layers.
Fig. 4 shows our results for the tip
located either on the highest or on the lowest
position on the reconstructed surface
(points $1$ and $2$ of Fig. 3a). Two
important conclusions can be drawn from our
results: first, the injected current along the metal 
layers is damped by the introduction of an
imaginary component for the energy, $E+i \eta$,
that simulates the electron-electron scattering
processes (in our case we have used 
$\eta=0.05$eV, that yields attenuation lengths 
in accordance with experimental data \cite{lee});
only at long distances this damping
results in an exponential behaviour for the
current. Second, at short distances the current
presents a faster decrease associated with
the injection of electrons along surface
states channels. As Fig. 4 shows, in
the $(2 \times 1)$ reconstruction the
current decreases by $65\%$ 
after the electrons cross the first two Si-layers and the 
first CoSi$_2$ trilayer (where surface states are 
mainly localized).
This is the effect of having the injected
electrons propagating also along the surface  
bands, departing in 
this way from the bulk states channels that
contribute to the current propagating across 
the metal layers.

\begin{figure}
\epsfxsize=0.5\textwidth \epsfbox{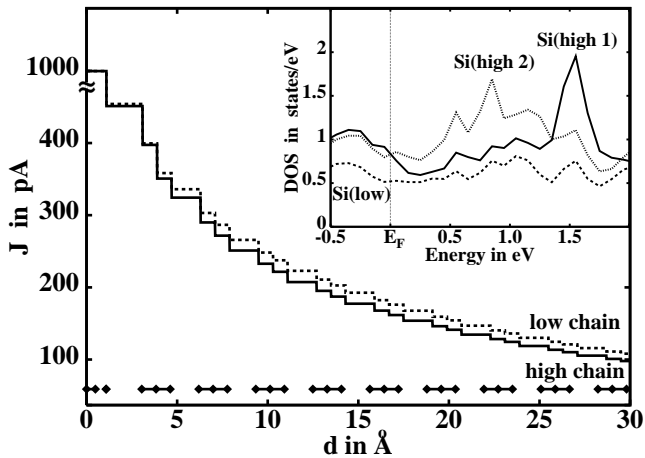}
\caption{
Current across the first 30{\AA} of the silicide film at $1.5$eV
after injection on a high chain (solid line)
or on a low chain (dotted line). The diamonds at the
bottom indicate the trilayer Si1-Co-Si2 sequence of CoSi${}_2$ in the
(111) direction, with the final reconstructed Si chain-bilayer.
The inset shows the surface density of states projected on the high
and low chain atoms in the energy region important for BEEM.} 
\label{damping}
\end{figure}

The anticorrugation obtained
in Fig. 3b for the total current arriving
at the M-S interface can be understood in terms of this
role played by the surface states.
These states have larger weights on the 
atoms of the higher chains than on the low 
chain ones as we can see in the surface 
density of states shown in the inset of Fig. 4. 
Therefore, for those tip positions
in which electrons are injected predominantly into
high chain atoms, there is a larger
probability for having those electrons 
propagating along surface states channels than
the one obtained for injection on the
low chain atoms. Consequently, in the usual experimental
constant-current STM mode, less current
crossing the metal layers and reaching the interface
remains. This difference with respect to the current
injection, which stays almost constant after crossing the 
first CoSi$_2$ trilayer (5{\AA}), explains why the  
BEEM image shows anticorrugation with 
respect to the surface topography.
Note, however, that the absolute order of magnitude 
of this effect depends strongly with energy: as shown 
in Fig. 3b, the anticorrugation contrast is $25\%$ for 
1.5 eV and a lower contrast is obtained for larger voltages.
This dependence is related to the fact that 
surface states are concentrated rather close to the 
Fermi level.

In conclusion, we have presented a theoretical analysis
of the propagation of an electron beam injected in a
CoSi$_{2}$(111) crystal using a STM tip. Our results show
conclusively that the silicide electronic band structure plays a
central role in the focalization of the electron beam.
This behaviour and the specific ${\bf k}_{\parallel}$-contribution
to the current have been associated with flat terraces
of the constant energy surface producing a {\it condenser lens} effect
on the electron propagation. Our results explain the
high resolution observed in real space for
BEEM experiments performed on CoSi$_{2}$/Si(111) interfaces.
Additionally, we have also shown how the BEEM current can map out 
the silicide surface reconstruction due to the role 
played by the localized CoSi$_{2}$(111)-(2x1) surface states
on the current injected from the tip.

We acknowledge financial support from 
the Spanish CICYT under
contracts number PB97-1224 and PB92-0168C.
K.R. and K.H. are grateful for financial support from
SFB292 (Germany).

\end{document}